\input harvmac

\lref\Eyn{
  B.~Eynard and N.~Orantin,
  ``Invariants of algebraic curves and topological expansion,''
  [arXiv:math-ph/0702045].
}
\lref\mar{
  B.~Eynard, M.~Marino and N.~Orantin,
  ``Holomorphic anomaly and matrix models,''
  JHEP {\bf 0706}, 058 (2007)
  [arXiv:hep-th/0702110].
}
\lref\Bouchard{
  V.~Bouchard, A.~Klemm, M.~Marino and S.~Pasquetti,
  ``Remodeling the B-model,''
  [arXiv:hep-th/07091453].
}
\lref\AganagicDB{
  M.~Aganagic, A.~Klemm, M.~Marino and C.~Vafa,
  ``The topological vertex,''
  Commun.\ Math.\ Phys.\  {\bf 254}, 425 (2005)
  [arXiv:hep-th/0305132].
}
\lref\DijkgraafFC{
  R.~Dijkgraaf and C.~Vafa,
  ``Matrix models, topological strings, and supersymmetric gauge theories,''
  Nucl.\ Phys.\  B {\bf 644}, 3 (2002)
  [arXiv:hep-th/0206255].
}

\lref\bcov{
  M.~Bershadsky, S.~Cecotti, H.~Ooguri and C.~Vafa,
  ``Kodaira-Spencer theory of gravity and exact results for quantum string
  amplitudes,''
  Commun.\ Math.\ Phys.\  {\bf 165}, 311 (1994)
  [arXiv:hep-th/9309140].
}

\lref\wit{
  E.~Witten,
  ``Quantum background independence in string theory,''
  [arXiv:hep-th/9306122].
}
\lref\dvv{
  R.~Dijkgraaf, E.~P.~Verlinde and M.~Vonk,
  ``On the partition sum of the NS five-brane,''
  [arXiv:hep-th/0205281].
}
\lref\dijko{
  R.~Dijkgraaf, E.~P.~Verlinde and H.~L.~Verlinde,
  ``C = 1 Conformal Field Theories on Riemann Surfaces,''
  Commun.\ Math.\ Phys.\  {\bf 115}, 649 (1988).
}
\lref\witcs{
  E.~Witten,
  ``Quantum field theory and the Jones polynomial,''
  Commun.\ Math.\ Phys.\  {\bf 121}, 351 (1989).
}

\lref\DijkgraafTE{
  R.~Dijkgraaf, S.~Gukov, A.~Neitzke and C.~Vafa,
  ``Topological M-theory as unification of form theories of gravity,''
  Adv.\ Theor.\ Math.\ Phys.\  {\bf 9}, 603 (2005)
  [arXiv:hep-th/0411073].
}

\lref\nek{
  N.~Nekrasov,
  ``Z-theory: Chasing M-F-Theory,''
  Comptes Rendus Physique {\bf 6}, 261 (2005).
}

\lref\WittenHC{
  E.~Witten,
  ``(2+1)-Dimensional Gravity as an Exactly Soluble System,''
  Nucl.\ Phys.\  B {\bf 311}, 46 (1988).
  }

\lref\wakimoto{
  M.~Wakimoto,
  ``Fock representations of the affine lie algebra A1(1),''
  Commun.\ Math.\ Phys.\  {\bf 104}, 605 (1986).
}

\lref\kpz{
  V.~G.~Knizhnik, A.~M.~Polyakov and A.~B.~Zamolodchikov,
  ``Fractal structure of 2d-quantum gravity,''
  Mod.\ Phys.\ Lett.\  A {\bf 3}, 819 (1988).
}

\lref\frenkel{
E. Frenkel, ``Lectures on Wakimoto modules, opers and the center at the
critical level,'' [arXiv:math.QA/0210029].  }

\lref\cubic{
  G.~T.~Horowitz, J.~D.~Lykken, R.~Rohm and A.~Strominger,
  ``A Purely Cubic Action For String Field Theory,''
  Phys.\ Rev.\ Lett.\  {\bf 57}, 283 (1986).
}

\lref\dijkchiral{
  R.~Dijkgraaf,
  ``Chiral deformations of conformal field theories,''
  Nucl.\ Phys.\  B {\bf 493}, 588 (1997)
  [arXiv:hep-th/9609022].
}

\def\figin{\epsfcheck\figin}\def\figins{\epsfcheck\figins}
\def\epsfcheck{\ifx\epsfbox\UnDeFiNeD
\message{(NO epsf.tex, FIGURES WILL BE IGNORED)}
\gdef\figin##1{\vskip2in}\gdef\figins##1{\hskip.5in}
\else\message{(FIGURES WILL BE INCLUDED)}%
\gdef\figin##1{##1}\gdef\figins##1{##1}\fi}
\def\DefWarn#1{}
\def\figinsert{\goodbreak\topinsert}
\def\ifig#1#2#3#4{\DefWarn#1\xdef#1{fig.~\the\figno}
\writedef{#1\leftbracket fig.\noexpand~\the\figno}%
\figinsert\figin{\centerline{\epsfxsize=#3mm \epsfbox{#2}}}
\bigskip\medskip\centerline{\vbox{\baselineskip12pt
\advance\hsize by -1truein\noindent\footnotefont{\sl
tFig.~\the\figno:}\sl\ #4}}
\bigskip\endinsert\noindent\global\advance\figno by1}

\def\dbar{{\overline{\partial}}}

\def\zbar{{\overline{z}}}

\def\C{{\bf C}}
\def\R{{\bf R}}
\def\Z{{\bf Z}}

\def\a{\alpha}

\def\l{\lambda}

\def\w{\omega}

\def\d{\partial}
\def\dbar{{\overline\partial}}

\def\cO{{\cal O}}

\def\cF{{\cal F}}

\def\({\bigl(}
\def\){\bigr)}
\def\<{\langle\,}
\def\>{\,\rangle}

\def\]{\right]}
\def\[{\left[}

\def\hb{\hat{\beta}}
\def\hg{\hat{\gamma}}

\Title
{\vbox{
\hbox{ITFA-2007-48}
}}
{\vbox{
\centerline{Two Dimensional Kodaira-Spencer Theory}
\vskip4mm
\centerline{and}
\vskip4mm
\centerline{Three Dimensional Chern-Simons Gravity}
}}
\centerline{Robbert Dijkgraaf}
\vskip.05in
\centerline{\sl Institute for Theoretical Physics \&
Korteweg-de Vries Institute for Mathematics}
\centerline{\sl  University of Amsterdam, Valckenierstraat
  65, 1018 XE Amsterdam, The Netherlands}
\medskip
\centerline{and}
\medskip
\centerline{Cumrun Vafa} \vskip.05in 
\centerline{\sl Jefferson Physical Laboratory, Harvard University, 
Cambridge, MA 02138, USA}
\smallskip

\vskip .5in
\centerline{\bf Abstract}

Motivated by the six-dimensional formulation of Kodaira-Spencer theory
for Calabi-Yau threefolds, we formulate a two-dimensional version and
argue that this is the relevant field theory for the target space of
local topological B-model with a geometry based on a Riemann surface.
We show that the Ward identities of this quantum theory is equivalent
to recursion relations recently proposed by Eynard and Orantin to
solve the topological B model.  Our derivation provides a conceptual
explanation of this link and reveals a hidden affine $SL_2$ symmetry.
Moreover we argue that our results provide the strongest evidence yet
of the existence of topological M theory in one higher dimension,
which in this case can be closely related to $SL_2$ Chern-Simons
formulation of three dimensional gravity.


\Date{November 2007}

\newsec{Introduction}

Topological strings have been solved in the context of local toric
Calabi-Yau threefolds.  In particular the topological vertex can be
used to compute all genus amplitudes for topological A-model on these
spaces \AganagicDB .  On the other hand, using mirror symmetry, this
construction can be interpreted as providing a full solution to the
B-model topological string with a local Calabi-Yau geometry modelled
on a Riemann surface (which we will refer to as the local B model).

There is however a more direct path to obtaining topological strings
in the context of the local B model: Matrix models are conjectured to
be equivalent to the topological B model on a local
geometry \DijkgraafFC, where the Riemann surface is identified as the
spectral curve of the matrix model.  This gives another solution to
the local B-models, namely the large $N$ 't Hooft expansion of the
corresponding matrix models.  There has been recent spectacular
progress in solving these matrix models where it has been shown that
the large $N$ description of matrix model can be directly formulated
{\it intrinsically} on the Riemann surface in terms of certain
recursion relations that essentially follow from the loop
equations \Eyn.  This new approach has the advantage that it applies
to any local B-model, whether or not the spectral curve comes from a
matrix model.  This relation has been recently checked in the context
a number of examples \Bouchard .  In the context of the B model the
approach of \Eyn\ has the remarkable feature that it automatically
incorporates the holomorphic anomaly \mar : The partition function
depends on the choice of A-cycles on the Riemann surface, and choosing
different basis for A-cycles leads to a generalized Fourier transform
of the partition function, as is expected on the basis of the general
holomorphic anomaly equation of the topological string. Usually this
fact is formulated as that the topological string partition function
transforms as a wave function or a holomorphic block.

The aim of this note is to derive the recursion relations of \Eyn\
directly in the B model using field theory techniques.  We will
demonstrate that these recursion relations are given in terms of Ward
identities of the B-model field theory, which is the restriction of
the Kodaira-Spencer theory on the Riemann surface.  Quite
surprisingly, while proving these recursion relations, we uncover an
$SL(2,\R)$ current algebra.  In this setup, the fact that the
partition function becomes a wave function is directly related to the
fact that one is dealing with a {\it chiral} boson on the Riemann
surface as the basic field of the gravity, and it is known that this
partition function does depend on the choice of a basis for the
A-cycles.  In fact the best understanding of this phenomenon comes
from the interpretation of the wave function as a state for the three
dimensional Chern-Simons theory.  Here we also speculate about the
existence of a topological M-theory, whose restriction to the local
case suggests the existence of a three dimensional gravity theory,
which leads to the topological strings as a quantum state.  This could
potentially explain the appearance of $SL(2,\R)$ current algebra on
the Riemann surface.

The organization of this paper is as follows: In section 2 we describe
the basic setup and formulate the relevant 2d KS theory.  In section 3
we show that the Ward identities of this theory are the same as those
written in \Eyn , and uncover an $SL(2,\R)$ symmetry.  In section 4 we
speculate about embedding this symmetry in one higher dimension.

\newsec{The Basic Setup}

Kodaira-Spencer theory is the string field theory of the topological
B-model on Calabi-Yau threefolds \bcov .  This theory can
be considered as the quantization of the $\dbar$ operator on the
Calabi-Yau manifold with a fixed complex structure, as captured by a
holomorphic $(3,0)$ form $\Omega$.  More precisely, it is the
quantization of the cohomologically {\it trivial} variations of
$\dbar$, which do not change a fixed background complex structure.
Thus the theory is defined in terms of a pair $(\dbar, \Omega)$.  It
is not known, at the present, how to use this formalism to solve all
genus amplitudes for compact Calabi-Yau manifolds.  One can compute,
however, low genus amplitudes using this approach.

As we will see, the situation is much better for the local non-compact
threefold modeled on a curve $\Sigma$.  In the context of these local
geometries it is natural to look for a reduction of this structure to
the Riemann surface $\Sigma$ and directly quantize that system. This
is the approach that we will follow in this note.

By the local case we shall mean a non-compact Calabi-Yau threefold
defined by the hypersurface:
$$
vw=H(x,y),
$$
where $v,w$ belong to ${\bf C}$ and $x,y$ belong to ${\bf C}$ or ${\bf
C}^*$. (To be precise, in the latter case the appropriate coordinates
are $e^x,e^y \in \C^*$ or, equivalently, $x,y \in \C/2\pi i \Z$.)  In
these coordinates the holomorphic three-form $\Omega$ is given by
$$
\Omega={dv\over v} \wedge dy \wedge dx.
$$
The local Riemann surface $\Sigma$ is defined by the equation
$$
H(x,y)=0.
$$
It is not difficult to see that the periods of the $(3,0)$ form
$\Omega$ on the local threefold can be reduced, upon integration to
the $x$-$y$ plane, to the integral of the one-form
$$
\omega = ydx
$$
on the Riemann surface.  We thus wish to define the quantum
Kodaira-Spencer theory of the pair $(\dbar, \omega)$, on the Riemann
surface $\Sigma$ given by $H(x,y)=0$.  More precisely, we wish to
integrate over all deformation of $\dbar$ which do not affect the
cohomology class of $\omega$, just as was the case for the 3-fold
case.

The variation of the complex structure is captured locally by the
deformation
$$
\dbar\rightarrow \dbar-\mu\partial,
$$
where the Beltrami differential $\mu$ is a tensor of type
$$
\mu = \mu_{\zbar}{}^z \,d\zbar \otimes \partial_z.
$$
For the variation to be globally trivial, it means that there is a 
diffeomorphism by a vector field $v=v^z\partial_z$ such that
$$
\mu_{\overline z}{}^z = \partial_\zbar v^z.
$$
We are interested in quantizing these deformations, while maintaining
the cohomology class of $\omega$.  As such, it is natural to formulate
the variation of the $\overline \partial$ in terms of its action on
$\omega$.  The condition of not changing the cohomology class of
$\omega=\omega_z dz$ means that
$$
\delta \omega = d\phi,
$$
for some function $\phi$. We will use the scalar $\phi$ as the basic
field of our KS theory.  In fact, we can re-express the vector field
$v^z$ in terms of $\phi$ as follows:
$$
v^z={\phi\over \omega_z}.
$$
To see this, note that the Lie derivative of $\omega$ in terms of 
$v$ can be expressed as
$$
\delta \omega={\cal L}_v\omega
=d(\iota_v\omega) -\iota_v d\omega =d(\iota_v \omega),
$$
since $\omega$ is closed. Note also that $\iota_v \omega=\omega \cdot
(\phi/ \omega)= \phi$.  This leads to the required relation
$\delta \omega= d\phi.$

In terms of the scalar field $\phi$ the variation of the $\dbar$
operator takes the form (using that $\dbar v=\dbar \phi/ \omega$)
$$
\dbar\rightarrow \dbar-{\dbar\phi \over \omega}\partial.
$$

Now before deformation the scalar field $\phi$ should satisfy
$$
\partial \dbar \phi =0.
$$
This one can see for example by making more explicit the dictionary
from the general KS theory on the threefold with the reduction to the
Riemann surface $\Sigma$. The KS field $A$ (the Beltrami differential)
of \bcov\ is identified as
$$
A \sim  {\dbar\phi \over \omega}. 
$$
The dual form $A'=\Omega \cdot A$, which is a $(2,1)$ form in six
dimensions, becomes in two dimensions a $(0,1)$ form given by
$$
A' \sim \dbar\phi.
$$
Now the closed string field $A'$ satisfies the gauge condition
$b_0^-A'=0$. In the KS theory this becomes $\d A'=0$. With the above
dictionary, this translates into $\d\dbar \phi=0$. (Closely related to
this point of view, we can also think of $\omega$ as the classical
value of $\d\phi$. Since this is a holomorphic (1,0) form, we have
(again classically) the equation $\dbar\d\phi=0$.)

This means that, at the level of the unperturbed equations, we are
dealing with a free (chiral) boson quantum field theory with action
(we will not be precise with normalizations)
$$
S = \int_\Sigma \partial \phi \dbar \phi.
$$

We can now capture the effect of the variation of $\dbar$ in terms of
an operator.  Recall that for any Beltrami differential
$\mu_{\zbar}{}^z$, the operator $T(\mu)$ which implements this
variation on the conformal field theory is given in terms of the
holomorphic stress-tensor $T_{zz}$ as
$$T(\mu)=\int_{\Sigma} T_{zz} \mu_{\zbar}{}^z.$$
Given that we have a free boson system, we can write this very
explicitly as
$$
T_{zz} ={1\over 2} \partial \phi \partial \phi.
$$
Using the fact that $\mu =\dbar\phi /\omega$, we therefore have the
interaction term
$$
\int_\Sigma \partial \phi \partial \phi {{\overline \partial 
\phi}\over \omega}.
$$
Note that this operator can be written as total derivative
$$
\int d\left(\partial \phi \partial \phi {\phi\over \omega}\right),
$$
where we use that in perturbation theory $\d\phi$ remains holomorphic.
So, if $\omega$ has no zeroes, this interaction is trivial.  Since for
our case $\omega= ydx$, such zeroes can occur if either $y=0$ or $dx=0$.  As
we will show in the next section, the points where $y=0$ do not
contribute, but the locus where $dx=0$ does.  The points where $dx=0$
correspond to branch points of the Riemann surface $H(x,y)=0$ on the
$x$-plane.  We will thus arrive at the interaction operator
$$
\sum_{branch\ points} \oint_{P}\partial \phi \partial \phi \cdot 
{\phi\over \omega},
$$
which will be used in the next section to recover the recursion
relations of \Eyn.

In fact there is one additional term in the action that we will now
explain: Note that the action we have thus far can be written as
$$
S=\int \partial \phi(\dbar +{\dbar 
\phi\over \omega}\partial )\phi
$$
However, as we have explained $d\phi$ is the variation of $\omega$.
In particular $\omega$, being a differential of type $(1,0)$, can be
viewed as the {\it classical} vev of $\partial \phi$.  Motivated by
this observation we view the first term in the above action as the
full $\omega$ including the classical piece and arrive at the final
form for the action
\eqn\action{
S=\int (\omega +\partial \phi)(\dbar +{\dbar \phi\over 
\omega}\partial )\phi
}
Expanding this action, and introducing the topological string coupling
constant $\l$ through the usual rescaling $\omega \to \omega/\l$ (a
standard relation in KS theory), we obtain the field theory
\eqn\pertaction{
S = \int \left[\d\phi \dbar\phi + {1\over \l}\omega \dbar\phi +
{\l\over \omega} \dbar\phi (\d\phi)^2 \right].
}
The first term is the usual kinetic term. The second term can be
interpreted as the coupling to a background holomorphic gauge field
$A_z = \omega/\l$. Since in perturbation theory the field $\phi$ is
chiral, this term will only influence the classical free energy, that
scales as $\l^{-2}$. Finally, the third term is capturing the
perturbative corrections. (In fact, by shifting $\del\phi$ with the
background $\omega$ the whole action can be put into cubic form,
reminiscent of the purely cubic forms encounter in open string field
theory \cubic .) 

In this action the cubic term is proportional to the string coupling
$\l$. So, up to several subtleties related to the chiral nature of
this quantum field theory, this model can be solved using trivalent
Feynman diagrams. In the next section we will show how one can use
this action to derive the recursion relations of \Eyn.

\newsec{The Recursion Relations}

We will now discuss the quantum field theory based on the
action \action\ in terms of coordinate space perturbation theory.
Following \Eyn\ we will consider not just the partition function, but
general correlation function of operators $\d\phi(z)$
$$
W(z_1,\ldots,z_n;\l)
= \Bigl\langle \d\phi(z_1) \cdots \d\phi(z_n)\Bigr\rangle_{con}.
$$
Here the subscript indicates that we only consider the connected correlators.
We compute these
correlators in the background of the interaction term
$$
\exp \int_\Sigma \l {\dbar\phi(\d\phi)^2 \over \omega}
$$
Expanding in the coupling $\l$ brings down these interactions and 
this defines the perturbative correlators. 
The connected correlators have an expansion of the form
$$
{W}(z_1,\ldots,z_n;\l) =
\sum_{g\geq 0} \l^{2g-2+n} {W}_g(z_1,\ldots,z_n).
$$

\subsec{The localisation to branch points}

As noted in the previous section, the interaction can be written as a
total derivative away from the zeroes of $\omega$. So we are left with
contributions of the form
\eqn\res{
\sum_P \oint_P {\phi(\d\phi)^2 \over \omega}
}
where $P \in \Sigma$ denote the positions of the zeroes of
$\omega$. In the local coordinates the one-form $\omega$ is given by
$\omega =ydx$, and such zeroes therefore occur either if $y(x)=0$, or
if the differential $dx$ vanishes. Let us consider these two cases
separately.

If $y(x)$ has a zero at $x=x_0$, so that $y \sim c (x - x_0)$, the
variable $z=x-x_0$ is a good local coordinate around this special
point and we can expand the quantum field as
$$
\d\phi(z) = \sum_{n \in \Z} \a_n z^{-n-1}.
$$
Here $\a_n$ are the usual creation and annihilation
operators. Plugging this relation and $\omega \sim zdz$ into
interaction vertex \res\ we obtain the operator
$$
\cO = \oint{dz \over z} \phi(\d\phi)^2 \sim 
\sum_{k+m+n=-1} {1\over k}\a_k \a_m \a_n
$$
Since $z$ is good local coordinate at this point, the field $\phi(z)$
has no singularities. In the operator formalism the operator $\cO$
should therefore be regarded to act on the vacuum $|0\rangle$. This
state satisfies $\a_n|0\rangle=0$ for $n \geq 0$. Because of the
condition that $k+m+n=-1$, we see that necessarily the mode expansion
of $\cO$ will have to contain annihilation operators $\a_{+n}$ that
vanish on this vacuum state. Therefore we obtain the relation
$$
\cO|0\rangle=0,
$$
and the action of the interaction at these zeroes is trivial and can
be ignored.

As we mentioned, a second source of zeroes of $\omega=ydx$ are the
points where $dx$ vanishes. If we think of the curve $H(x,y)=0$ as an
orbit in the phase space of Hamiltonian mechanics, these are the
turning points. In complex geometry these zeroes are branch points of
the algebraic curve. Generically, these are simple branch points. At
such a point the curve is locally described by
$$
(y-y_0)^2 = x-x_0.
$$
A good local coordinate is therefore
$$
z=y-y_0 = (x-x_0)^{1/2},
$$
where we clearly see that we are dealing with a branch point. In terms
of the variable $z$ we have $dx \sim zdz$ and therefore also
$\omega \sim zdz$ (since $y$ attains the regular value $y_0$ at this
point). So $\omega$ has indeed a (single) zero at the ramification
point $z=0$. Now in terms of the coordinate $x$ the interaction vertex
$\cO$ does {\it not} act not on the regular vacuum state $|0\rangle$,
but on a {\it twisted} state $|\sigma\rangle$, around which the field
$\d\phi(x)$ has a half-integer mode expansion
$$
\d\phi(x) \sim \sum_{n\in \Z} \a_{n-{1\over 2}}(x-x_0)^{-n-{1\over
2}}.
$$
Equivalently, in terms of ``good'' coordinate $z$ on the double cover
the scalar field satisfies the condition
\eqn\twist{
\phi(-z) = -\phi(z).
}
This behaviour can be understood from the fact that the perturbed
one-form $\omega + d\phi$ should have the same behaviour as $\omega$,
which around the ramification point takes the form $zdz$. This fixes
the boundary condition on the field $\phi(z)$ and determines in turn
the nature of the boundary state $|\sigma\rangle$. Working out the
decomposition of $\cO$ in these twisted modes we easily see that in
this case $\cO|\sigma\rangle \not = 0$. Therefore the branch points
give non-vanishing contributions to the interaction and we are left
with
\eqn\branch{
\sum_{branch\ points} \oint_P {\phi(\d\phi)^2 \over \omega}
}

We will now have to contract the insertions $\d\phi(z_i)$ with the
operators appearing in the interaction vertex. If $z$ is the local
coordinate around the branch point $P$ this gives terms of the form
$$
\left\langle \d\phi(w) \oint_z 
{\phi(z) \d\phi(z) \d\phi(z) \over \omega(z)} \cdots \right\rangle
$$

To evaluate these kinds of expressions we need to determine the chiral
correlator
$$
B(z,w) = \left\langle \d\phi(z) \d\phi(w)\right\rangle
$$
for a free boson on the surface $\Sigma$.  This two-point function is
well-know to be given by the Bergmann kernel. To define it uniquely,
we have to fix the loop momenta through a set of $A$-cycles
$$
\oint_{A_I} \d \phi = p_I.
$$
The standard kernel $B(z,w)$ takes these $p_I=0$. Note that this
prescription already breaks the modular invariance, since $Sp(2g,\Z)$
transformations, that relate one set of homology cycles to another,
will act via generalized Fourier transformation on the correlators of
the chiral boson.

Now first, up to total derivatives, we can consider the contraction of
$\d\phi(w)$ with $\phi(z)$. This will be given by the primitive of the
Bergmann kernel that we will denote as $G(z,w)$
$$
\langle \phi(z) \d\phi(w) \rangle = G(z,w) := \int^z \!\!B(v,w) dv.
$$
However we should also take into account the boundary
condition \twist\ at the branch point. The field $\d\phi$ should be
anti-periodic around the twist field insertion. This we can enforce by
inserting an explicit projector to the odd part of the propagator, as
is customary in the computation of twist field correlation functions
in orbifold models. So we get (in the local coordinate $z$, close to
the branch point)
$$
\langle \phi(z) \d\phi(w) \rangle_{twist}  = {1\over 2} \int_{-z}^z \!\!
B(v,w) dv.
$$
Note in particular that at the branch point $z=0$ this twisted
propagator vanishes, which is forced by the anti-periodicity. However,
in this case there is a matching zero in the denominator, because also
$\w$ vanishes at the branch point. We can therefore apply
l'H\^{o}pital's rule, and consider instead the limit
$$
\lim_{z \to 0} {\langle \phi(z) \d\phi(w) \rangle_{twist} \over \w(z)} 
= \lim_{z \to 0} { \int_{\tilde{z}}^z B(v,w) dv \over \w(z)
 - \w(\tilde{z})}
$$
where $z$ and $\tilde{z}$ are the two points on the two branches of
$\Sigma$ that project to the same image in the $x$-plane (so that
$\tilde{z} \sim -z$ close to $z=0$).

Finally we also have to deal with the fact that the interaction
consists of a cubic term that has to be normal ordered. Now recall
that this term originated from the stress-tensor insertion
\eqn\vT{
\oint v^z T_{zz},
}
where the vector field was taken to be $v^z=\phi/\omega$.  The
self-interactions in the stress-tensor are usually defined by
point-splitting regularization
$$
T(z) = \lim_{\tilde{z} \to z}{1\over
2} \left[ \d\phi(z)\d\phi(\tilde{z}) - {1 \over
(z-\tilde{z})^2}\right]
$$
To have a consist perturbation theory we have to insert this
definition into equation \vT\ with $v^z=\phi/\omega$.

Summarizing all this, and up to some further subtleties that we
discuss in section 3.3, we obtain the recursion relation of \Eyn
. This relation can be considered as the Schwinger-Dyson equations for
the interacting boson field theory. Expanding the recursion relation
gives a graphical representation in terms of trivalent Feynman
diagrams. There will be both connected and disconnected contributions.
For the connected diagrams with $n+1$ external legs the recursion
relation takes the form
\eqn\eynrec{
\eqalign{
{W}_g(w,z_1,z_2,\ldots,z_n) =  & \sum_{P} {\rm Res}_{\strut z=P} 
 { \int_{\tilde{z}}^z B(v,w)
 dv \over \w(z)- \w(\tilde{z})}\Bigl[ 
{W}_{g-1}(z,\tilde{z},z_2,\ldots,z_n)\cr
 & \ \ \ + \sum_{h=0}^g \sum_{Z = Z' \cup
 Z''}{W}_h(z,z'_1,\ldots,z'_m) 
{W}_{g-h}(\tilde{z},z''_1,\ldots,z''_{n-m})
\Bigr] 
}}
Here $P$ is summed over all branch points of the spectral curve
$\Sigma$. The set $Z$ denotes the collection of ``free'' marked points
$Z=\{z_1,\ldots,z_n\}$, and in the disconnected piece there is a sum
over all splittings of the set $Z$ into two disjoint (and possibly
empty) subsets $Z'$ and $Z''$ of order $m$ and $n-m$.

\subsec{The partition function}

We now turn to the partition function itself, which has an expansion
$$
Z = \exp \cF, \qquad \cF = \sum_{g\geq 0} \l^{2g-2} \cF_g.
$$
To derive the final recursion relation of \Eyn\ we will employ a
rescaling symmetry. Starting point will be again the complete action
\pertaction\ that we recall here for convenience
$$
S = \int \d\phi \dbar\phi + {1\over \l}\omega \dbar\phi +
{\l\over \omega} (\d\phi)^2\dbar\phi.
$$
Consider now the action of the vector field $\l \d_\l$ on the free
energy $\cF$. On the one hand we clearly have
\eqn\fg{ 
\l {\d \cF \over \d\l} = \sum_{g\geq 0} (2g-2) \cF_g \l^{2g-2}.
}
On the other hand the effect of this rescaling on the action has the
effect of bringing down two possible terms.  One of them is the cubic
interaction term we have focused thus far. In this case of the vacuum
amplitude, following the logic of the previous derivation, this term
does not contribute, as there are no other $\partial \phi$
observables inserted anywhere.  Only the second term in the action
will contribute and gives the insertion
$$
\l{\d Z \over \d \l} =
-{1\over \lambda} \bigl\langle \int_\Sigma \omega \dbar\phi \bigr\rangle.
$$
Just as for the cubic interaction term, it is convenient to write this
as a total derivative ${1\over \lambda}\int d(\omega \phi)$, which
gives zero, except when $\phi$ has poles.  This happens when $\phi$ is
near the branch points $P$.  So we can write this term as
$$
{1\over \lambda}\sum_{branch\
points} \bigl\langle\oint_{P}\omega \phi \bigr\rangle.
$$
Near the branch points we write $\omega = \d\phi_{cl}$, where
$\phi_{cl}(z)$ can be interpreted as the classical value of the field
$\phi$ and so this contour term can be written (using integration by
parts) as
\eqn\onept{
-{1\over \l} \sum_{P} \bigl\langle \oint_P \phi_{cl} \d\phi \bigr\rangle
=
-{1\over \l} \sum_{P} \oint \phi_{cl} \bigl\langle \d\phi \bigr\rangle
=
-{1\over \l} \sum_P {\rm Res}_{\strut z=P}\phi_{cl}(z)W(z)
}
Now the normalized (or connected) one-point function
${W}(z)$ has a perturbative expansion
$$
{W}(z) = \sum_{g\geq 0} \l^{2g-1} {W}_g(z)
$$
Inserting this expansion into \onept\ and comparing this with \fg\ we
thus find the recursion relation of \Eyn
$$ 
\cF_g={1\over 2-2g}\sum_P {\rm 
Res}_{\strut z=P}\phi_{cl}(z){W}_g(z).
$$

\subsec{The chiral projection and a hidden affine $SL(2,\R)$ symmetry}

Up to this point the relation of the action \pertaction\ to the
manipulations leading to the recursion relations has not been very
precise. Indeed the authors of \Eyn\ have remarked that their
recursion relations and the corresponding graphical solution cannot
follow from a straightforward Feynman expansion, since certain
contractions and diagrams are missing, and there is a special order in
which the vertices are connected. This last point is due to the fact
that the interactions have been rewritten as contour integrals, which
we can think of as a (local) Hamiltonian formalism. These operators in
general do not commute, and the specific time ordering prescription
breaks the general covariance. Exactly the same point was met
in \dijkchiral .

From our point of view another source of subtleties arise because we
are dealing with a chiral field theory and up to now we have not
consistently implemented the chiral projection on the scalar field
$\phi$. As written in action \pertaction\ the anti-holomorphic
component of $\phi$ is a propagating field. This gives unwanted
contributions in the contractions of $\phi$ with itself. We will now
rectify this point.

At the level of the free theory the projection onto the chiral modes
can be done by adding a multiplier (1,0) form $\gamma$ to the action
$$ 
\int \left( {1\over2} \del\phi\dbar\phi - \gamma \dbar \phi\right).
$$
Integrating out $\gamma$ enforces the chiral condition
$\dbar\phi=0$. On the other hand, integrating out $\phi$ expresses
$\gamma$ as
$$
\gamma = \del\phi.
$$
In fact, it is now suggestive to relabel
$$
\phi=\beta,
$$
to make clear that we are dealing with a bosonic $\beta$-$\gamma$
system with spins 0 and 1. After a partial integration, this
$\beta\gamma$ system has the action
$$
\int_\Sigma \left({1\over 2} \del\beta\dbar\beta + 
\beta\dbar \gamma\right) + \oint_\infty \beta\gamma,
$$
where by $\oint_\infty$ here and in the following we mean integration
over the boundaries of the Riemann surface if there are any (including
the branch points).  With this unconventional action we get the
following Green's functions for the fields $\beta$ and $\gamma$
\eqn\props{
\eqalign{
\langle \beta(z) \beta(w) \rangle   & = 0,\cr
\langle \beta(z) \gamma(w) \rangle  & = G(z,w),\cr
\langle \gamma(z) \gamma(w) \rangle & = B(z,w).\cr
}}
Here, as before, $\d_zG(z,w)=B(z,w)$.

Using this formalism we can now write the interaction term in an
elegant fashion as
\eqn\ointeg{
\oint_\infty \left({\omega \over \lambda} \beta + \beta\gamma +
{\lambda\over \omega} \beta \gamma^2\right).
}
Here the middle term is just added for suggestive reasons; it does not
contribute as long as the total momentum flowing in or out of the
boundary of $\Sigma$ is zero.

The main advantage of writing the action like this is that it
automatically reproduces the diagrammatics of \Eyn. The correlators
that are considered are of the form
$$
\bigl\langle \gamma(z_1) \ldots \gamma(z_n) \bigr\rangle.
$$
The string loop interactions now come from the $\beta\gamma^2$ term.
Since $\beta$ only contracts with $\gamma$
each  $\beta$ in the interaction term $\beta\gamma^2$ gets
paired up with a $\gamma(z_i)$ leaving two additional $\gamma$'s.
At the end when all the $\beta$'s have been contracted, we use the 
$\gamma$ correlations and thus recover the \Eyn\ recursion relations
as well as the boundary conditions needed to solve them.
This reformulation makes our derivation of the recursion relation precise.

Interestingly, this way of expressing the action suggests a hidden
$SL(2,\R)$ symmetry.  Let us first recall the Wakimoto representation
of the $SL(2,\R)_k$ current algebra \wakimoto. This conformal field
theory consists of another $\beta\gamma$ system, that we will write
as $(\hb,\hg)$ and that now has spins 1 and 0, together with an extra
scalar field $\chi$. In terms of these variables the $SL(2,\R)$
currents, all of spin 1 of course, are expressed as
\eqn\sltwo{
\eqalign{
J_+(z) & = \hb, \cr
J_3(z) & = \hb\hg + {1\over 2}\a_+ \d\chi, \cr
J_-(z) & = \hb\hg^2 + \a_+ \hg\d\chi + k\d\hg.\cr
}}
Here $\a_+^2=2k-4$ and the central charge is given by $c={3k\over
k-2}$. The scalar field $\chi$ has a background charge
$1/\a_+$. Furthermore, in order to compute correlation functions one
needs to add various insertions of the screening charge
\eqn\screen{
S_+ = \oint_\infty \hb \,e^{-2\chi/\a_+}.
}

In the application we have in mind $\chi$ does not appear and thus it
is natural to view this as the special limit of $k=2$. In this
so-called critical limit, where the central charge $c\to \infty$ and
$\a_+ \to 0$, the contribution of the scalar $\chi$ decouples and can
be ignored. In fact the $\beta$-$\gamma$ system by itself carries a
representation of the current algebra $SL(2,\R)$ at level $k=2$. (In
the analytic continuation to the case of $SU(2)$ this critical level
corresponds to the value $k=-2$.) However, this representation is far
from irreducible. For example, there is a very large center spanned by
the modes of the limiting case of the rescaled stress-tensor
$$
u(z) = (k-2)T(z) = \sum_a J_a^2,
$$
(instead of just the identity operator). One way to describe the
representations in this critical limit, is that the combination
$v(z)=\a_+ \del\chi(z)$ becomes a classical, non-dynamical scalar
field that parametrizes the affine $SL(2,\R)$ representations. This
classical scalar field can be traded with the expectation values of
the rescaled stress-tensor $u(z)$. All of this is intimitely connected
to the theory of integrable systems, see \frenkel\ for more on this.

There is a traditional topological twisting of this model related to
the KPZ model of 2d gravity \kpz, where the spins of the
$\beta$-$\gamma$ system are changed from $(1,0)$ to $(0,1)$. In this
twist the triplet of affine $SL(2,\R)$ currents $(J_+,J_3,J_-)$ change
its spins into $(0,1,2)$. This twisting needs a section of the
canonical bundle $K_\Sigma$ on the Riemann surface, {\it i.e.}, we
have to pick a meromorphic (1,0) form $\omega$. The twisted fields
$(\beta,\gamma)$ are related to the untwisted fields as
$$
\hat\beta = \omega \beta,\qquad \hat\gamma = {1\over \omega} \gamma.
$$
Ignoring the scalar $\chi$ and total derivatives we so find that the
triplet of currents can be expressed as
\eqn\sltwist{
\eqalign{
J_+(z) & = \omega \beta, \cr
J_3(z) & = \beta\gamma, \cr
J_-(z) & = {1\over \omega} \beta\gamma^2.\cr
}}
Using this notation we can take the interaction of the KS theory, that
we have managed to put in the form
$$
\oint_\infty \left({\omega \over \lambda} \beta + \beta\gamma +
{\lambda\over \omega} \beta \gamma^2\right),
$$
and rewrite it in a more suggestive way as
\eqn\Jplus{
\oint_\infty J_+(z,\l), 
}
where we introduced a one-parameter family of currents
$$
J_+(z,\l) = {1\over \l}J_+(z) + J_3(z) + \l J_-(z).
$$
Here the variable $\l$ can be seen as a spectral parameter that picks
a $U(1)$ inside $SL(2,R)$, or more geometrically a null plane inside
$\R^{2,1}$. It therefore takes it values on the twistor ``sphere'', or
perhaps more correctly the twistor upper-half plane (the appropriate
real structure is not quite obvious)
$$
\l \in {\bf H} = {SL(2,R) \over U(1)}.
$$
The group $SL(2,R)$ acts on this parameter in the usual fashion by
fractional linear transformations
$$
\l \to {a\l + b \over c \l + d},\qquad ad-bc=1.
$$
Possibly one can think of the interaction \Jplus\ as a (singular)
$k=2$ limit of the screening charge $S_+$ given in \screen. Anyway,
there is in the critical case $k=2$ a huge algebra that commutes with
the interaction vertex $J_+(\l)$, namely the full classical Virasoro
algebra. Itis not clear what the precise role of this structure is and
toi which extend it is related to other models of two-dimensional
gravity. In the next section we discuss the potential meaning of the
$SL(2,\R)$ structure from a three-dimensional perspective.

\newsec{The 3d Gravity Lift}

An important aspect of the formulation of topological B model in the
setup of \Eyn\ is the fact that it in a natural way explains \mar\ why
the partition function satisfies the holomorphic anomaly equation
of \bcov .  In particular to even formulate the partition function,
one has to choose a set of A-cycles on the Riemann surface, such that
the period of the Bergmann kernel vanishes over them.  If we chose a
different set of A-cycles, the partition function transformorms as a
wave function as is the interpretation of holomorphic anomaly
equation \wit , see also \dvv .

In our setup this comes about because we have a chiral boson $\phi$
and to define it quantum mechanically we need to specify its periods
over only half the cycles, which in the present case are the A-cycles,
where $\int_{A_i} \partial \phi =0$.  It is known from the study of 2d
CFT's \dijko\ that this is indeed consistent with how the chiral
blocks are defined, and how they transform if we choose a different
symplectic basis for them.  For example, the chiral blocks transform
according to Fourier transform, if we switch the A-cycles and
B-cycles.  This phenomenon for 2d CFT's is best described by
considering the associated three dimensinonal Chern-Simons
theory \witcs: Consider the abelian $U(1)$ Chern-Simons theory in
three dimension, with the action
$$\int A \wedge dA$$
where $A$ denotes the $U(1)$ connection.  In particular in the Coulomb
gauge we have the term $\int A_x \dot{A}_y$, which means $A_x$ and
$A_y$ are conjugate variables.  Viewing the three dimensional manifold
as Riemann surface times time, and identifying the chiral fields of 2d
with the restriction of $A=d\phi$ on the Riemann surface, we see the
fact that the periods of $d\phi$ over the $A$-cycles and $B$-cycles
form conjugate variables and do not commute.

It is very natural to view this as the natural motivation for us as
well.  For the general topological string this suggests the existence
of a topological M-theory in one higher dimension, namely 7 for the
Calabi-Yau threefold, or 3 for the current case (where the CY is based
on a Riemann surface).  This in fact was one of the motivations for
the introduction of topological M theory \refs{\DijkgraafTE, \nek} .

However the abelian Chern-Simons theory in 3d cannot be the whole
story here for two reasons: First of all we have in addition to the
free term the interactions in 2d, and they should lead to some
deformations of the 3d theory. Secondly the 2d theory is a {\it
gravity} theory, and not a current algebra, and so we expect that the
3d theory to also be a gravity theory. It is natural to ask whether we
can relate our theory to the $SL(2,\R)$ Chern-Simons formulation of 3d
gravity \WittenHC\ in the geometry $\Sigma \times \R$.  Such a 3d
theory would give rise to chiral $SL(2,\R)$ current algebra living on
$\Sigma$.  But this is surprisingly what we have found in the last
section! 

We are thus led to believe that 3d gravitational Chern-Simons theory
underlies what we have found in connection with the Kodaira-Spencer
theory on the Riemann surface.  However we need to better understand
the meaning of the insertion terms
$$
\oint_\infty J_+(z,\l).
$$
First of all, what is the twistorial meaning of $\l$?  In other words,
why should the coupling constant of topological string parameterize a
twistorial sphere and what is the role of the $SL_2$ transformations
acting on the coupling constant? Here it should be stressed that the
relevant object is the local form $\omega/\l$, which can be viewed as
a varying point on the twistor sphere. This choice of background is
what distinguishes the 3d theory from ordinary (chiral) $SL(2,\R)$
gravity. 

Secondly, why would one insert this interaction term at the boundaries
and branch points of the surface?  Is it related to the screening
operators at the critical level?  One interpretation may be that this
term shifts the gravitational background so that the connection is not
flat and would correspond to the background $\Sigma \times \R$.  If
this is the right interpretation one would need to better understand
why this particular insertion creates this gravitational background.

Recalling that we have been discussing only the reduction of
Kodaira-Spencer theory to two dimensions, we could lift our findings
back to six dimensions.  In this context we would be led to the seven
dimensional M-theory formulation, mentioned above.  We feel we have
found a first concrete evidence for the existence of topological
M-theory.  It would be very important to deepen our understanding of
the 3d lift of Kodaira-Spencer theory, and further extend it to the
topological M-theory in seven dimensions.

\vglue 2 cm

\noindent {\bf Acknowledgements}

We would like to thank J. de Boer, V. Bouchard, B. Eynard, A. Klemm,
H. Ooguri and E. Verlinde for valuable discussions.  We would also
like to thank the fifth Simons Workshop in mathematics and physics at
Stony Brook for inspiring surroundings and discussions which led to
this project.  The research of R.D. was supported by a NWO Spinoza
grant and the FOM program {\it String Theory and Quantum Gravity}.
The research of C.V. was supported in part by NSF grants PHY-0244821
and DMS-0244464.

\listrefs

\end